\documentclass[aps,pre,twocolumn,showpacs,floatfix,superscriptaddress, graphics]{revtex4-1}
\usepackage{amsmath,amssymb,graphicx,color,braket,hyphenat,xcolor}
\usepackage[ocgcolorlinks,colorlinks=true,linkcolor=blue,citecolor=red,linktocpage=true]{hyperref}

\newcommand{\tr}{\mathrm{Tr}}

\begin{document}
\title{Boosting the performance of small autonomous refrigerators \\ via common environmental effects}
\author{Gonzalo Manzano}
\affiliation{International Centre for Theoretical Physics ICTP, Strada Costiera 11, 4151 Trieste, Italy}
\affiliation{Scuola Normale Superiore, Piazza dei Cavalieri 7, I-56126, Pisa, Italy}

\author{Gian-Luca Giorgi}
\affiliation{Institute for Cross-disciplinary Physics and Complex Systems, IFISC (UIB-CSIC), Campus Universitat de les Illes Balears, E-07122 Palma de Mallorca, Spain}

\author{Rosario Fazio}
\affiliation{International Centre for Theoretical Physics ICTP, Strada Costiera 11, 4151 Trieste, Italy}
\affiliation{Dipartimento di Fisica, Universit\'a di Napoli ``Federico II'', Monte S. Angelo, I-80126 Napoli, Italy}

\author{Roberta Zambrini}
\affiliation{Institute for Cross-disciplinary Physics and Complex Systems, IFISC (UIB-CSIC), Campus Universitat de les Illes Balears, E-07122 Palma de Mallorca, Spain}
\begin{abstract}
We explore the possibility of enhancing the performance of small thermal machines by the presence of common noise sources. In particular, we study a prototypical model for an autonomous quantum refrigerator comprised by three qubits coupled to thermal reservoirs at different temperatures. Our results show that engineering the coupling to the reservoirs to act as common environments lead to relevant improvements in the performance. 
The enhancements arrive to almost double the cooling power of the original fridge without compromising its efficiency. 
The greater enhancements are obtained when the refrigerator may benefit from the presence of a decoherence-free subspace. The influence of coherent effects in the dissipation due to one- and two-spin correlated processes is also examined by comparison with an equivalent incoherent yet correlated model of dissipation.
\end{abstract}
\maketitle
\section{Introduction}
\label{intro}
The characterization of the performance of quantum thermal machines constitutes an important objective of quantum thermodynamics, an emerging field at the intersection of quantum information science and nonequilibrium thermodynamics~\cite{Goold,Anders,Beretta}. 
These devices consist of a small quantum system able to complete some beneficial thermodynamic task, such as work extraction, refrigeration, pumping heat, etc. 
Since the first works exploring three-level masers as heat engines or refrigerators \cite{Scovil, Geusic, Geusic2}, a plethora of different models operating either in cycles \cite{QuanCycles, Campisi, KosloffOtto} or in a continuous fashion \cite{Kosloffcontinuous, Gelbwaser, Uzdin} have been proposed. 
They provide both theoretical insights and proposals for implementations in the laboratory, some of which are recently seeing light \cite{Pekola, LutzIon, Mottonen, AdsorptionIons, Poem}. 
Furthermore, quantum thermal machines may be of practical importance in biological processes \cite{Scully, Killoran}, for measuring time \cite{Huber} or small temperatures \cite{Hofer}, as well as producing quantum resources such as entanglement \cite{Brask} or coherence \cite{Manzano}.

A particularly interesting class of quantum thermal machines are autonomous quantum refrigerators~\cite{Popescu, Paul, BrunnerVirtual, Entanglement, Silva-Manzano}, also called absorption refrigerators~\cite{Palao, Levy, Alonso, Adesso, SilvaR, Segal, Seah, Latune}. 
They represent simple models comprised by few qubits (or harmonic oscillators) or a single qutrit weakly interacting with thermal reservoirs at different temperatures. They are self-contained configurations, 
thus avoiding external sources of coherence and control, which may incur in non-trivial thermodynamic costs~\cite{Huber, Woods} (for a comparison see Ref.~\cite{Clivaz, Clivaz2}). 
Proposals for their implementation in the laboratory include quantum-dot models~\cite{Qdots,Qdots2}, a setup with ions in optical cavities~\cite{Mitchison} or in circuit QED architectures~\cite{ChenQED, QED-fridge}. 
The first experimental realization of a quantum autonomous refrigerator has been recently reported using trapped ions~\cite{AdsorptionIons}.

The performance of autonomous quantum refrigerators has been extensively studied in most prototypical configurations~\cite{Popescu, Paul, BrunnerVirtual, Levy, Alonso, Adesso}, in some of which the presence of steady-state entanglement~\cite{Entanglement} and quantum discord~\cite{Alonso} has been found for particular regimes of parameters. 
Nevertheless, in these refrigerators, the ability to cool is limited by design constraints like the Hilbert space dimension of the machine~\cite{Silva-Manzano} or the maximum energy gap in the machine Hamiltonian~\cite{Clivaz, Clivaz2}, and to be surpassed extra resources are needed. 
In particular, it has been shown that when replacing one of the thermal reservoirs by a squeezed thermal reservoir (or other non-thermal reservoirs) both the cooling power and the efficiency can be greatly improved~\cite{Adesso,Latune} (see also Ref.~\cite{MaxFridge}).

Here we pursue a different way to improve the performance of autonomous quantum refrigerators, namely, allowing the thermal reservoirs to act as common reservoirs over the whole machine~\cite{PekolaCB}. 
This implies that all the energy transitions in the machine with a given energy gap, will couple to the same environment, now inducing correlated transitions over them. 
The possibility of common or collective dissipation can be traced all the way back to the Dicke model and the discovery of superradiance~\cite{Dicke}. 
Common reservoirs have been also traditionally considered in decoherence and dissipation models for quantum computers~\cite{Palma, Zanardi}. 
Nowadays it is well known that they may lead to decoherence free subspaces and subsystems~\cite{LidarR} enabling strategies for bypassing decoherence in quantum registers~\cite{NQCodes, Pairing}, 
as well as for entanglement generation and preservation~\cite{Braun02, Benatti03, Aguado, Paz, Zell, ManzanoSync, ManCB}. 

Collective dissipation can be conveniently engineered in the laboratory or naturally arise in the presence of isotropic environments if the systems are sufficiently close to each others~\cite{Dicke,Palma,Zell}. 
Furthermore, in structured environments collective dissipation can arise even between distant bodies~\cite{Galve,Tudela}.

We propose an autonomous refrigerator model comprising just three qubits coupled to common thermal reservoirs that, contrary to previous approaches based on Otto cycles in superconducting qubits~\cite{PekolaCB, PekolaOtto}, it is able to benefit from collective dissipation for improving its performance. 
In particular we will see that engineering common noise sources over the refrigerator level structure, it can be enhanced up to a point almost doubling the cooling power reached for local dissipation. 
Moreover, we show that these enhancements are improved in the presence of decoherence free subspaces, then providing a genuine quantum enhancement of the refrigerator performance.

\section{Three qubit refrigerator model}
\label{sec:setup}

\begin{figure}
\includegraphics[width= \linewidth]{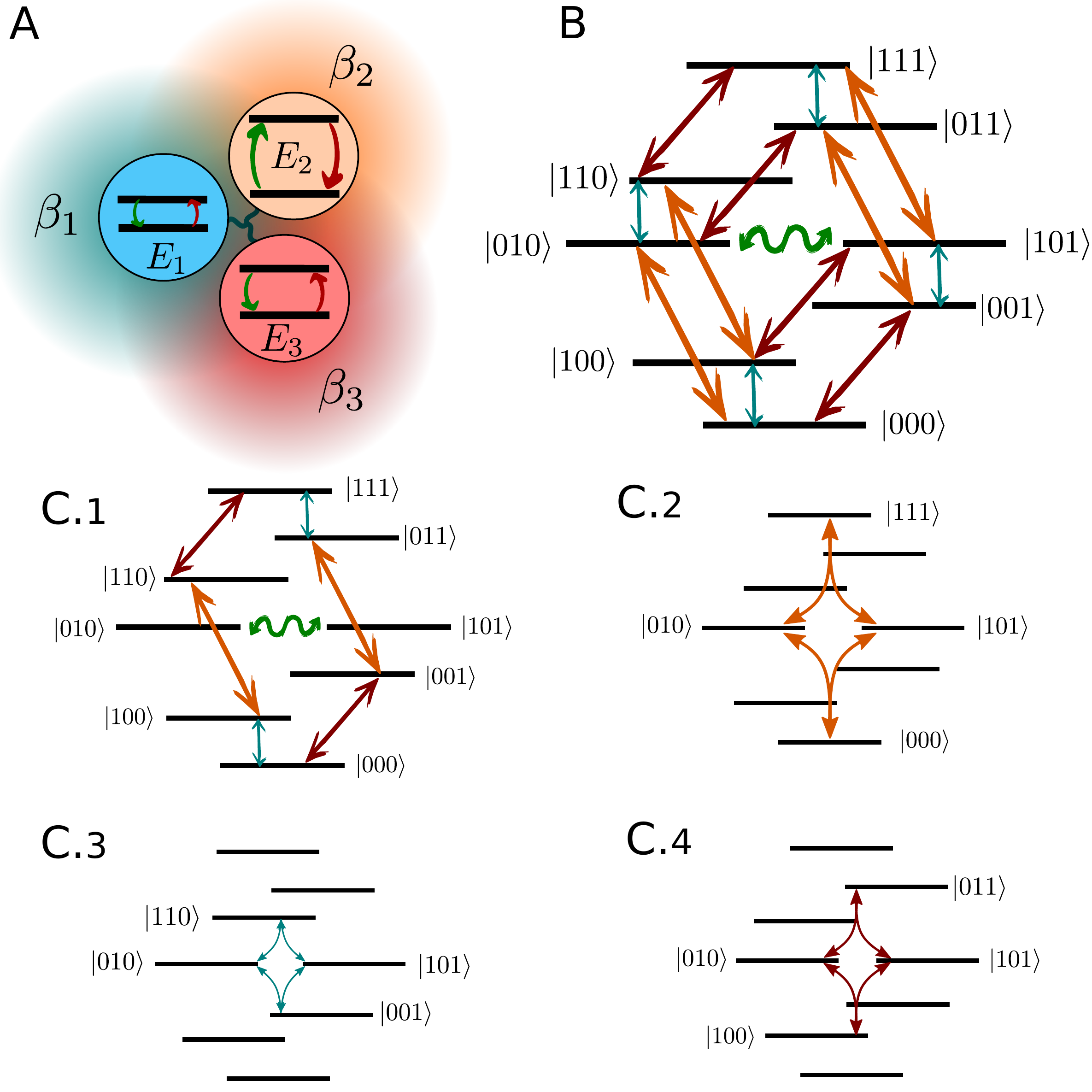}
\caption{(A) Schematic representation of the three qubit refrigerator model coupled to three thermal reservoirs at different inverse temperatures $\beta_1$, $\beta_2$ and $\beta_3$. 
(B) The three-body Hamiltonian of the refrigerator together with the transitions between energy levels promoted by the three different reservoirs in the original configuration with separate baths. 
The green arrow stands for the coherent exchange induced by the refrigerator interaction Hamiltonian $H_\mathrm{int}$. (C) For the case of common reservoirs we divided the induced transitions in four sets (C.1 - C.4) all of which act in parallel. Some of the previous transitions  from (to) the degenerate levels now become coherent transitions, as represented by the double arrows in (C.2 - C.4), inducing jumps between single levels and the superposition state $(\ket{010} + \ket{101})/\sqrt{2}$. \label{fig:1}}
\end{figure}

We focus on the refrigerator originally introduced in Refs. \cite{Popescu, Paul}. In this model the machine is comprised by three qubits with different energy spacings $E_1$, $E_2$, and $E_3$, constrained by 
the relation $E_3 = E_2 - E_1$ (see Fig. \ref{fig:1}). The Hamiltonian reads
\begin{equation}
 H_\mathrm{m} = H_1 + H_2 + H_3 + H_\mathrm{int},
\end{equation}
with $H_i = E_i \ket{1}\bra{1}_i$ for $i=1,2,3$ the Hamiltonian of each qubit (we employ the computational basis of each qubit $\{ \ket{0}_i, \ket{1}_i\}$), and a three-body weak interaction between them
\begin{equation}\label{eq:int}
 H_\mathrm{int} = g \left( \ket{101}\bra{010} + \ket{010}\bra{101} \right),
\end{equation}
with $g \ll E_i$. This energy-preserving interaction induces transitions between the degenerate levels $\ket{010}$ and $\ket{101}$ without requiring any input work, that is $[H_\mathrm{m}, H_\mathrm{int}] = 0$. 

The idea underlying the functioning of the refrigerator can be summarized as follows. Assume that the three qubits of the machine are in thermal equilibrium at the same temperature, $\rho_\mathrm{m} = \rho_1^\beta \otimes \rho_2^\beta \otimes \rho_3^\beta$ with $\rho_i^{\beta} = e^{-\beta E_i}/Z_i$ the thermal (Gibbs) states for each qubit, $Z_i = \tr[e^{-\beta E_i}]$ being the partition function and $\beta = 1/ k_B T$ the inverse temperature. Then the populations of the states $\ket{101}$ and $\ket{010}$ will be equal due to the constraint $E_3 = E_2 - E_1$. Consequently the interaction $H_\mathrm{int}$ in Eq.~\eqref{eq:int} will swap back and forth between states $\ket{101} \leftrightarrow \ket{010}$ with the same probability in any direction. Nevertheless, as soon as the qubits are allowed to be at different temperatures, the population of one of the two degenerate states may be favored, and the transitions can be biased. For instance one may cool qubit $1$ by promoting transitions $\ket{101} \rightarrow \
\ket{010}$ over the backward ones, which is accomplished e.g. by assuming a sufficient high temperature in qubit $3$ with respect to qubits $1$ and $2$. Therefore, in order to obtain a refrigerator showing a continuous mode of operation we couple each qubit of the machine to a different thermal reservoir at temperatures $T_1 < T_2 < T_3$, or equivalently $\beta_1 > \beta_2 > \beta_3$. 

\subsection{Three distinct common reservoirs}

In order to model the contact with thermal baths, the original works considering this setup used a phenomenological master equation based on a reset 
model~\cite{Popescu,Paul,BrunnerVirtual, Entanglement}. A microscopic derivation of the dynamics instead considers explicitly the weak interaction with the 
thermal reservoirs~\cite{Alonso, Adesso}, the latter being modeled as infinite sets of bosonic modes in thermal equilibrium~\cite{BreuerBook}. We follow this second approach.
However, in contrast to previous references, we consider that the three thermal reservoirs are common to the whole machine \and distinct. In particular,  
each of them will couple not only to a single qubit, but also to other energetic transitions in the machine with the same spacing. In this way, reservoir $i$ at inverse temperature $\beta_i$, will couple to qubit $i$ but also to the other transitions at $E_i$ present in the composed Hilbert space of the other two qubits (see the schematic representation of Fig.~\ref{fig:1}) enabled by the constraint $E_3 = E_2 - E_1$. The description of the machine-reservoirs interaction Hamiltonian is given in Appendix~\ref{appA}. This model may be of practical importance when the thermal reservoirs are coupled to the machine through frequency filters, but the qubits are not enough separated in physical space so that one cannot guarantee that each of them couples only to a single reservoir.


\subsection{Master equation in the local approach}

Assuming weak coupling between the machine and reservoirs and a rapid decay of the reservoirs correlation functions, a master equation in the Lindblad form \cite{Lindblad} can be derived within the standard Born-Markov approximation \cite{BreuerBook}.
Furthermore, as the coupling among the three qubits is assumed to be weak ($g\ll E_i$), a local approach can be taken such that the dissipative part of the master equation can be calculated neglecting the  inter-qubit coupling, which would only enter in the coherent part of the evolution \cite{RivasMEQ, HoferMEQ, GonzalezMEQ, PlenioMEQ,CattaneoMEQ}. 
It is worth mentioning that even if it has been argued that the local approach may lead to violations of the second law in a specific configuration~\cite{Violations}, these deviations are so small that fall below the order of magnitude employed to derive the master equation, and then should be simply neglected~\cite{Volovich}. 

Following the above considerations, we will model the dissipative dynamics of our setup within the local approach (more details are given in appendix \ref{appA}). We obtain the following master equation in the Schr\"odinger picture:
\begin{equation} \label{eq:master}
\dot{\rho}_\mathrm{m} = -\frac{i}{\hbar}[H_\mathrm{m}, \rho_\mathrm{m}] + \sum_i \mathcal{L}_i(\rho_\mathrm{m}), 
\end{equation}
with the three Lindbladians accounting for the dissipative effect of each reservoirs
\begin{align} \label{eq:lindbladians}
 \mathcal{L}_i(\rho_\mathrm{m}) &= \gamma_\downarrow^{i} \left(s_i \rho_\mathrm{m} s_i^\dagger - \frac{1}{2}\{ s_i^\dagger s_i, \rho_\mathrm{m} \} \right) \nonumber \\
& ~+ \gamma_\uparrow^{i} \left(s_i^\dagger \rho_\mathrm{m} s_i - \frac{1}{2}\{ s_i s_i^\dagger, \rho_\mathrm{m} \} \right).
 \end{align}
The relevant jump operators characterizing the dynamics are given by 
\begin{align} \label{eq:systemops}
 s_1 &= \sigma_1^{-} + \alpha \sigma_2^{-} \sigma_3^{+}, \nonumber \\ 
 s_2 &= \sigma_2^{-} + \alpha \sigma_1^{-} \sigma_3^{-}, \nonumber \\  
 s_3 &= \sigma_3^{-} + \alpha \sigma_1^{+} \sigma_2^{-},
\end{align}
where $\sigma_i^{-} = \ket{0}\bra{1}_i$, and the parameter $\alpha \in [0, 1]$ controls the degree of coupling of the reservoirs to the rest of the machine other than the single qubits. 
That is, for $\alpha = 0$, we recover the original model with three separate baths, while $\alpha = 1$ represents the case of three common reservoirs inducing completely correlated one- and two-spin transitions. 

Each jump operator \eqref{eq:systemops} represents the joint flip of one and two spins interacting with a bath mode at the same frequencyand introduces delocalized dissipative effects in the dynamics. 
This will play an important role linking the dynamics of diagonal (populations) and non-diagonal (coherences) elements of the machine density operator $\rho_\mathrm{m}$. 
Moreover in Eq.~\eqref{eq:lindbladians} we introduced the rates $\gamma_\downarrow^{(i)} \equiv \gamma_0^{(i)} (n_\mathrm{th}^{(i)} + 1)$ and $\gamma_\uparrow^{(i)} \equiv \gamma_0^{(i)} n_\mathrm{th}^{(i)}$, 
with $n_\mathrm{th}^{(i)}=(e^{\beta_i E_i} -1)^{-1}$ and $\gamma_0^{(i)}$ the spontaneous emission rates, 
fulfilling local detailed balance conditions $\gamma_\downarrow^{(i)} = \gamma_\uparrow^{(i)} e^{\beta_i E_i}$. 

A final remark on the derivation of Eq.~\eqref{eq:master} is in order. It is important  not to confuse the modeling of the thermal reservoirs as common or separate reservoirs with the local or global character of the master equation. 
While the first distinction arises from the type of coupling considered in the system-reservoir interaction the second distinguishes between the different approximations leading to a final form of the master equation which depends on the magnitude of the inter-system coupling. 
For example, the Lindbladians obtained in Eq.~\eqref{eq:lindbladians} with the operators in Eq.~\eqref{eq:systemops} are fundamentally different from the ones obtained for the case of separate reservoirs under the global approach~\cite{Alonso, Seah}. 

\subsection{Steady state regime}

We are mostly interested in the long term behavior of the refrigerator, so that in this paper we will restrict ourselves to the regime in which the machine reaches a steady state $\pi_\mathrm{m}$. This state verifies
\begin{equation}
  -\frac{i}{\hbar}[H_\mathrm{m}, \pi_\mathrm{m}] + \sum_i \mathcal{L}_i(\pi_\mathrm{m}) = 0.
\end{equation}
In order to obtain $\pi_\mathrm{m}$, we may proceed as follows. First, we derive from Eq.~\eqref{eq:master} a set of ten coupled differential equations linking the evolution of the populations of the eight energy levels of the 
refrigerators, $\{ p_{000}, p_{001}, ..., p_{111} \}$, with the real and imaginary parts of the coherence between the degenerate levels, $\{c_\mathcal{R}, c_\mathcal{I}\}$. They can be written in matrix form as
\begin{equation} \label{eq:evolution}
 \dot{\bold{p}} = \mathcal{W} ~\bold{p}, ~~\mathrm{with}~~~ \bold{p} = (p_{000}, ..., p_{111}, c_\mathcal{R}, c_\mathcal{I})^T,
\end{equation}
where $\mathcal{W}$ is a $10\times10$ matrix containing the full information about the dynamical evolution. On the other hand, the rest of coherences not appearing in Eq.~\eqref{eq:evolution} are quickly damped by dissipative effects and can be neglected in the long-time run. In particular, the equations governing the degenerate levels $\ket{101}$ an $\ket{010}$ together with its coherence read (the full set of equations is given in Appendix~\ref{appB}):
\begin{widetext} 
\begin{align} \label{eq:dynamics1}
 \dot{p}_{101} =&~ \gamma_\uparrow^{(2)} p_{000} + \alpha^2 \gamma_\uparrow^{(1)} p_{001} + \alpha^2 \gamma_\uparrow^{(3)}p_{100}  + \gamma_\downarrow^{(3)} p_{011} + \gamma_\downarrow^{(1)} p_{110} + \alpha^2 \gamma_\downarrow^{(2)} p_{111} - 2i g c_\mathcal{I} \nonumber \\
                &~  - \alpha \sum_i (\gamma_\uparrow^{(i)} + \gamma_\downarrow^{(i)}) c_\mathcal{R} -[\gamma^{(1)}_\uparrow + \gamma^{(2)}_\downarrow + \gamma^{(3)}_\uparrow + \alpha^2 (\gamma^{(1)}_\downarrow + \gamma^{(2)}_\uparrow + \gamma^{(3)}_\downarrow)] p_{010}, \\
 \dot{p}_{010} =&~ \alpha^2 \gamma_\uparrow^{(2)} p_{000} + \gamma_\uparrow^{(1)} p_{001} + \gamma_\uparrow^{(3)}p_{100}  + \alpha^2 \gamma_\downarrow^{(3)} p_{011} + \alpha^2 \gamma_\downarrow^{(1)} p_{110} + \gamma_\downarrow^{(2)} p_{111} + 2i g c_\mathcal{I} \nonumber \\
                &~  - \alpha \sum_i (\gamma_\uparrow^{(i)} + \gamma_\downarrow^{(i)}) c_\mathcal{R} -[\alpha^2 (\gamma^{(1)}_\uparrow + \gamma^{(2)}_\downarrow + \gamma^{(3)}_\uparrow) + \gamma^{(1)}_\downarrow + \gamma^{(2)}_\uparrow + \gamma^{(3)}_\downarrow] p_{010}, \\
 \label{eq:dynamics3}
 \dot{c}_\mathrm{R} =&~ \alpha (\gamma^{(2)}_\uparrow p_{000} + \gamma^{(1)}_\uparrow p_{001} + \gamma^{(3)}_\uparrow p_{100} + \gamma^{(3)}_\downarrow p_{011} + \gamma^{(1)}_\downarrow p_{110} + \gamma^{(2)}_\downarrow p_{111}) \nonumber \\
                     &~ - \alpha \sum_i (\gamma^{(i)}_\uparrow + \gamma^{(i)}_\downarrow)(p_{010} + p_{101}) - (\alpha^2 +1)\sum_i (\gamma^{(i)}_\uparrow + \gamma^{(i)}_\downarrow), 
 \end{align}
\end{widetext}  
and $\dot{c}_\mathcal{I} = i g (p_{101} - p_{010})$. Some insight can be gained by looking at the terms multiplied by $\alpha$ in the above equations. We notice that the effect of the common reservoirs is twofold: first they induce extra transitions between levels at resonant energy not present in the original model (terms with $\alpha^2$) as illustrated in Fig.~\ref{fig:1}. Second they introduce a link between populations and the real part of the coherence between degenerate levels (terms with $\alpha$) of quantum origin. 

Quantum effects in the evolution become dramatic in the case $\alpha = 1$, where we find a dark state (or decoherence-free subspace~\cite{Lidar}) of the dynamics
\begin{equation} \label{eq:dark}
 \ket{\psi_D} \equiv \frac{1}{\sqrt{2}} \left( \ket{010} - \ket{101} \right).
\end{equation}
This state is protected against dissipation and decoherence and therefore its population $p_D = \tr[\ket{\psi_D}\bra{\psi_D} \rho_\mathrm{m}]$ is conserved through the entire dynamical evolution. Indeed by noticing that $p_D = (p_{010} + p_{101})/2 - c_\mathrm{R}$, it can be easily checked from Eqs.~\eqref{eq:dynamics1}-\eqref{eq:dynamics3} above that  $\dot{p}_D = (\dot{p}_{010} + \dot{p}_{101})/2 - \dot{c}_\mathrm{R} = 0$. As we will see, this effect has interesting consequences in the performance of the machine. 

The steady state solution of the dynamics can be finally obtained from the eigenvectors of the transition matrix $\mathcal{W}$ with corresponding eigenvalues equal to zero. In the case $\alpha \in [0, 1)$ there is a single eigenvector with zero eigenvalue, then providing a unique steady state $\bold{\pi} = \{\pi_{000}, \pi_{001}, ..., \pi_{111}, c_\mathcal{R}^\pi, c_\mathcal{I}^\pi \}$. On the other hand, for $\alpha= 1$ the steady state will be sensible to the initial population of the dark state, $p_D$. Due to the dimension of the matrix $\mathcal{W}$ and the several parameters of the model, we obtain numerically the eigenvectors and eigenvalues leading to the steady state solutions $\bold{\pi}$. This is all we need to analyze the relevant thermodynamic quantities and present our main results in Sec. \ref{sec:results}.

\subsection{Characterization of the performance } 

Here we introduce the key thermodynamic quantities used to characterize the performance of the refrigerator. In first place, we are interested in the average heat currents from the reservoirs, $\dot{Q}_i$, and in particular 
in the heat current from the reservoir at the lower temperature, $\dot{Q}_1$, usually referred as the {\it cooling power}. The latter is a measure of the power of the refrigeration process when the reservoir at $\beta_1$ is considered the (macroscopic) 
object to be cooled. All three heat currents can be easily calculated from the master equation~\eqref{eq:master} as $\dot{Q}_i \equiv \tr[H_\mathrm{m} \mathcal{L}_i(\rho)]$ for $i= 1, 2, 3$. They read 
\begin{widetext} 
\begin{align} \label{eq:currents}
 \dot{Q}_1 &= E_1 \left( \gamma_\uparrow^{(1)} [\pi_0^{(1)} + \alpha^2(\pi_{001} + \pi_{101}) + 2 \alpha c_\mathcal{R}^\pi] - \gamma_\downarrow^{(1)}[\pi_1^{(1)} + \alpha^2(\pi_{110} + \pi_{010}) + 2 \alpha c_\mathcal{R}^\pi] \right), \\
 \dot{Q}_2 &= E_2 \left( \gamma_\uparrow^{(2)} [\pi_0^{(2)} + \alpha^2(\pi_{000} + \pi_{010}) + 2 \alpha c_\mathcal{R}^\pi] - \gamma_\downarrow^{(2)}[\pi_1^{(2)} + \alpha^2(\pi_{111} + \pi_{101}) + 2 \alpha c_\mathcal{R}^\pi] \right), \\ \label{eq:currents3}
 \dot{Q}_3 &= E_3 \left( \gamma_\uparrow^{(3)} [\pi_0^{(3)} + \alpha^2(\pi_{100} + \pi_{101}) + 2 \alpha c_\mathcal{R}^\pi] - \gamma_\downarrow^{(3)}[\pi_1^{(3)} + \alpha^2(\pi_{011} + \pi_{010}) + 2 \alpha c_\mathcal{R}^\pi] \right),
 \end{align}
 \end{widetext}
where we denoted for convenience $\pi_0^{(1)} = \sum_{k,l} \pi_{0kl}$, $\pi_1^{(1)} = \sum_{k,l} \pi_{1kl}$, and analogously for $\pi_0^{(2)}$, $\pi_1^{(2)}$,$\pi_0^{(3)}$ and $\pi_1^{(3)}$. Notice that all currents depend on the real part of the coherence between the degenerate levels $c_\mathrm{R}^\pi$. In the above expressions, the terms $\tr[H_\mathrm{int} \mathcal{L}_i(\pi)] \sim g \gamma_0^{(i)}$ have been neglected in accordance to the weak coupling limit employed in the derivation of the master equation~\eqref{eq:master}. Indeed since $g \ll E_i$, and $\gamma_0^{(i)}$ is second order in the system-reservoirs coupling, these terms correspond to higher order contributions to the heat currents and should be neglected. Moreover it is easy to check that the first law of thermodynamics in the steady state, $\dot{Q}_1 + \dot{Q}_2 + \dot{Q}_3 = 0$, is fulfilled.

The second law of thermodynamics in the steady state regime can be stated as the positivity of the total entropy production rate \cite{Kosloffcontinuous}
\begin{align} \label{eq:second}
\dot{\Sigma} = - \sum_i \beta_i \dot{Q}_i = \dot{Q}_3 (\beta_2 - \beta_3) - \dot{Q}_1 (\beta_1 - \beta_2) \geq 0,
\end{align}
where in the last equality we used the first law. The entropy production rate $\dot{\Sigma}$ measures the irreversibility in the operation of the refrigerator, and imposes the ultimate bounds on its efficiency. 

The efficiency of refrigeration is defined here by the ratio between the cooling power and the heat current from the hottest reservoir, i.e. the so-called coefficient of performance (COP) fulfilling
\begin{equation}\label{eq:carnot}
 \eta \equiv \frac{\dot{Q_1}}{\dot{Q}_3} \leq \frac{\beta_2 - \beta_3}{\beta_1 - \beta_2} \equiv \eta_C,
\end{equation}
where $\eta_C$ is the Carnot COP \cite{Kosloffcontinuous}, which can be regarded as the Carnot efficiency for refrigerators \cite{Paul}. Furthermore, as we will see, the heat currents in the model fulfill the general relation 
\begin{equation}\label{eq:window}
|\dot{Q}_i/ \dot{Q}_j| = E_i/E_j, ~~~ \forall i,j = 1, 2,3, 
\end{equation}
originally noticed in the qutrit model of Ref.~\cite{Geusic} (except for $\alpha = 1$, a case which we will treat later on a separate basis). This relation has been also reported for the present fridge model with separate reservoirs~\cite{Paul} 
and checked to break down in the case of higher qubits coupling $g$ as described by the global approach~\cite{Adesso}. 
When Eq.~\eqref{eq:window} is combined with Eq.~\eqref{eq:carnot}, we obtain a design constraint for the energy spacings of the machine qubits if they are to reach the refrigeration regime $(\dot{Q}_1 \geq 0)$:
\begin{equation}\label{eq:coolingwindow}
E_1 \leq \frac{\beta_2 - \beta_3}{\beta_1 - \beta_2} E_3  = \eta_C  E_3. 
\end{equation}
This has been called in the literature the {\it cooling window} \cite{Palao}. 

In the limit $E_1 \rightarrow \eta_C  E_3$, when $\eta \rightarrow \eta_C$, all heat current vanishes, 
and then Carnot COP can only be reached at exactly zero cooling power. Henceforth, another figure of merit to characterize the performance of the refrigerator is the COP at maximum cooling power, which we denote by $\eta_\ast$. 
This quantity has been extensively studied within the global approach in Ref. \cite{Alonso, Seah}, and shown to be bounded by a fraction of $\eta_C$ only depending on the dimensionality of the cold reservoir \cite{Alonso, Adesso}.

Finally, we are also interested in considering the qubit $1$ as the object to be cooled \cite{Popescu}. In such case the performance of the refrigeration process can be characterized by the effective local temperature 
of this qubit in the steady state \cite{Entanglement, Silva-Manzano, Clivaz}. The effective inverse temperature of qubit $i$ can be defined by means of the Gibbs ratio 
\begin{equation} \label{eq:efftemp}
\beta_i^\mathrm{eff} \equiv E_i^{-1} \ln \left( \pi_0^{(i)}/\pi_1^{(i)} \right).
\end{equation}
We notice that the interpretation of $\beta_i^\mathrm{eff}$ as an inverse temperature may become problematic in the case in which the reduced state of the corresponding qubit shows some amount of local coherence, but this is not the case for the steady state regime of the present model.

\section{Results}
\label{sec:results}

In the following we report our results on the performance enhancements in refrigeration due to the presence of common reservoirs, which are summarized in Figs. \ref{fig:2}, \ref{fig:3}, and \ref{fig:4}. 
Moreover in Fig. \ref{fig:5} we discuss the origin of the enhancements, identifying the effects of the coherent action of one- and two-spin dissipation processes.
To obtain the reported results, we calculate numerically the steady state of the fridge for different set of parameters and compute the different thermodynamic figures of merit introduced in the previous section. 
In order to provide a meaningful characterization we fix the inverse temperature of the reservoir at the coldest temperature, $\beta_1$, setting the reference energy scale by $k_B T_1$. 
Furthermore, we limit the largest energy gap on the machine qubits to $E_2 = 5 k_B T_1$, since otherwise the ability to cool of the fridge can be made arbitrarily large~\cite{Silva-Manzano}. 
Finally, for the ease of simplicity, we assume symmetric spontaneous emission rates $\gamma_0^{(i)} \equiv \gamma_0 = 0.01 k_B T_1$ for $i=1,2,3$.

\subsection{Enhanced cooling}
\label{sec:enhancements}

\begin{figure}[t!]
\includegraphics[width= \linewidth]{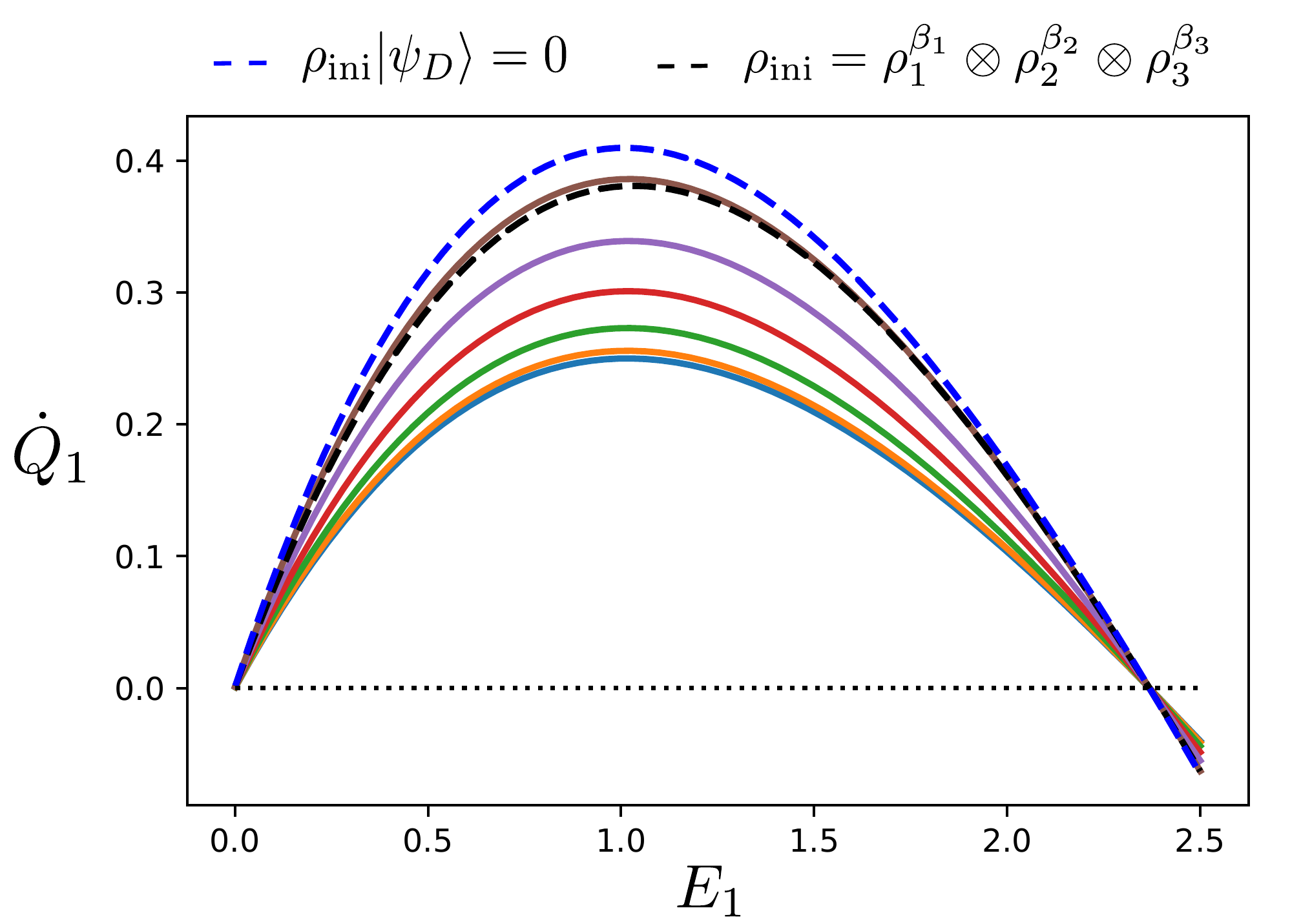}
\caption{Cooling power $\dot{Q}_1$ as a function of $E_1$ for different values of $\alpha = \{0, 0.2, 0.4, 0.6, 0.8, 0.99 \}$ (solid lines from bottom to top). Dashed lines corresponds to two different choices of the initial state in the case $\alpha = 1$. Temperatures of the reservoirs are $\beta_2 = 0.5 \beta_1$ and $\beta_3 = 0.05 \beta_1$, and we set $g= 0.005$.
Cooling power is given in $\gamma_0$ units.  \label{fig:2}}
\end{figure}

Our first main result is that for all the range of parameters explored, we obtain a significant enhancement in the cooling ability of the refrigerator (see Fig. \ref{fig:2}). These improvements can be characterized by looking at the cooling power, $\dot{Q}_1$ in Eq.~\eqref{eq:currents}, for different values of 
the parameter $\alpha$ controlling the common character of the reservoirs. 

In Fig. \ref{fig:2} we plot the cooling power as a function of the energy spacing $E_1$. Here we see that increasing $\alpha$ results in an increasing cooling power inside the cooling window. As anticipated in the previous section, we notice that 
the cooling window does not depend on $\alpha$. Special attention requires the case $\alpha = 1$ since in this case the cooling power depends on the initial state of the machine, $\rho_\mathrm{ini}$. In the plot the two dashed lines corresponds to two different choices of the initial state when $\alpha = 1$: the black dashed 
line is for an initial product state with the three qubits in equilibrium with their respective reservoirs, and the blue dashed one corresponds to an initial state orthogonal to the dark state, $\rho_{\mathrm{ini}} \ket{\psi_D} = 0$. As we can see the second initial state leads to an improved cooling power (we will turn back to this point later). 
In any case, these improvements in cooling power come at no extra cost in the efficiency of the refrigerator, whose COP is found to be nearly constant when increasing $\alpha$. 

\begin{figure*}[t!]
\includegraphics[width= 0.9 \linewidth]{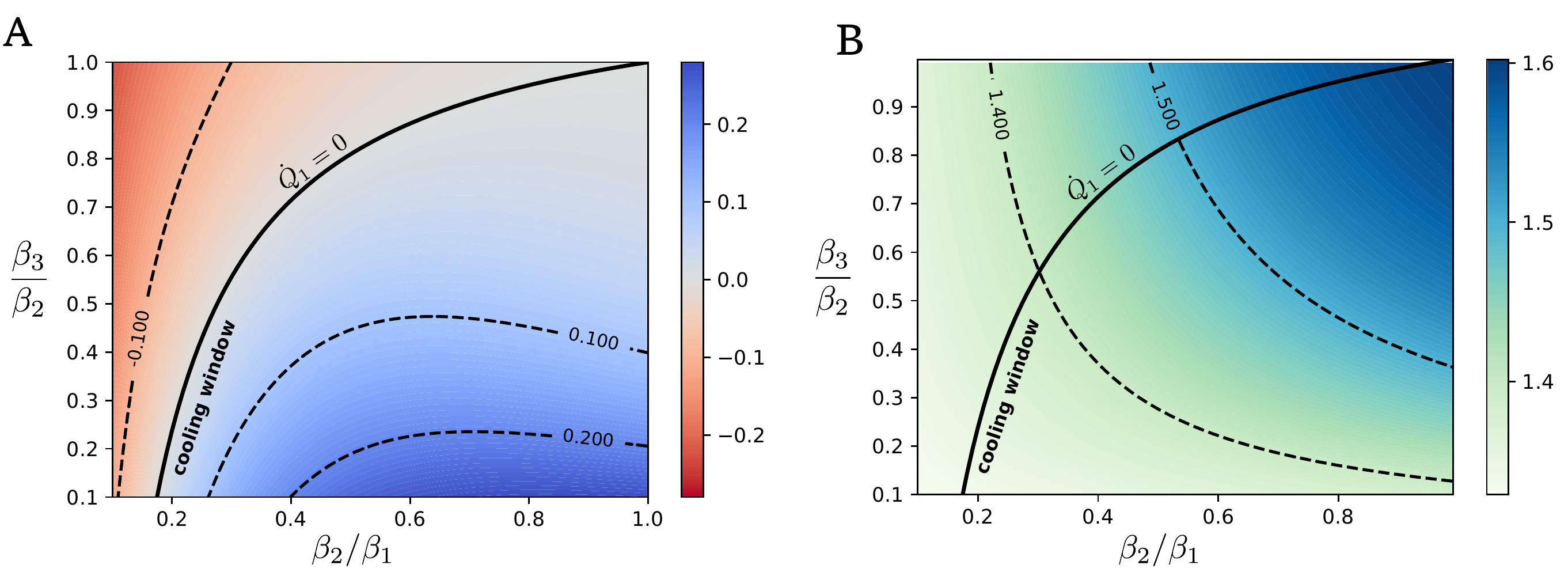}
\includegraphics[width= 0.9 \linewidth]{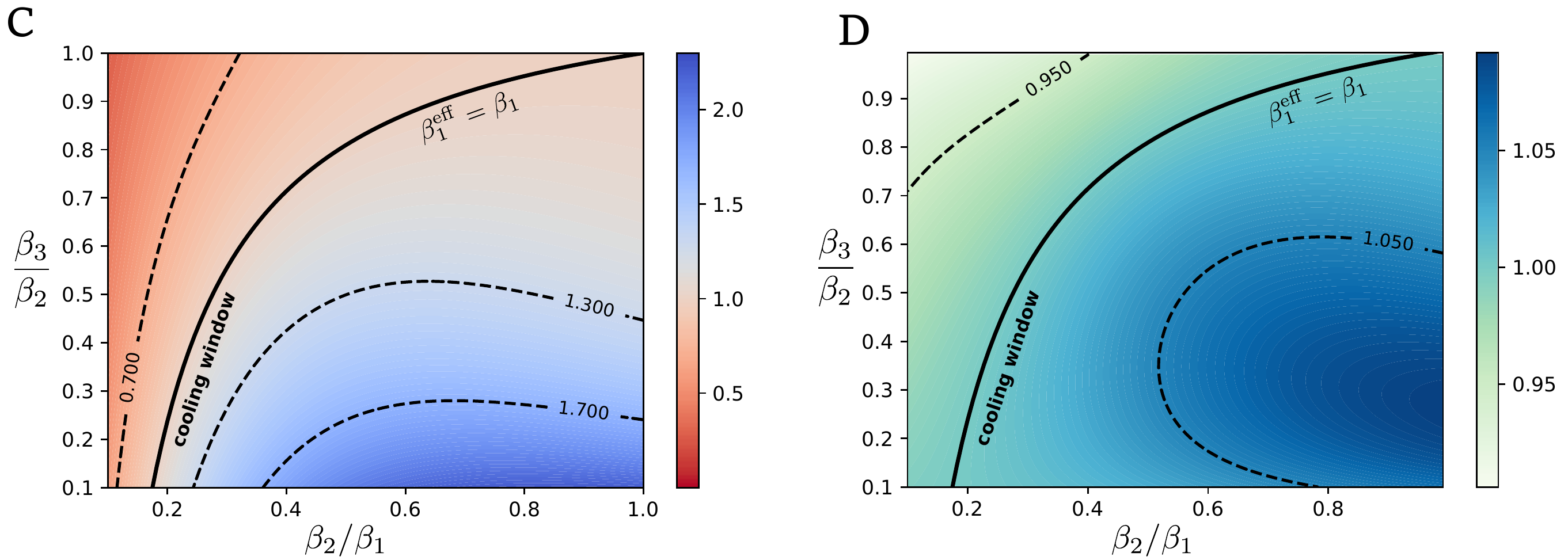}
\caption{(A) Cooling power $\dot{Q}_1$ as a function of $\beta_2/\beta_1$ and $\beta_3/\beta_2$ and (B) enhancements in the cooling power relative to the separate reservoirs case, $\dot{Q}_1/\dot{\bar{Q}}_1$, for $\alpha = 0.8$ and $E_1 = 0.8 k_B T_1$. 
Again cooling power is given in $\gamma_0$ units. (C) Effective temperature of qubit $1$, $\bar{\beta}_1^\mathrm{eff}$ as a function of $\beta_2$ and $\beta_3$, and (D) enhancements $\beta_1^\mathrm{eff}/\bar{\beta}_1^\mathrm{eff}$ for $\alpha = 0.8$. In all plots $g= 0. 01$. \label{fig:3}}
\end{figure*}

The dependence of the cooling power with the temperatures of the reservoirs is illustrated by the density plots in Fig. \ref{fig:3}. In Fig. \ref{fig:3}A we show the cooling power for purely local reservoirs, $\alpha = 0$, which we denote from now on $\dot{\bar{Q}}_1$, when the inverse temperatures of reservoirs are modified. Blue tones denote a higher cooling power while red ones denote heat dissipation into the coldest reservoir $1$. The black solid line corresponds to $\dot{Q}_1 =0$, which occurs when equality in Eq.~\eqref{eq:coolingwindow} is reached. As expected, the refrigerator stops running either for high temperatures of reservoir $2$ (left region) or when the temperature of reservoir $3$ is not high enough (top region). In any case $\beta_1 \geq \beta_2 \geq \beta_3$.

We obtain the same qualitative behavior when considering common reservoirs, $\alpha > 0$. However, now we find an amplification of heat fluxes over the very same regions of Fig. \ref{fig:3}A. In particular in Fig. \ref{fig:3}B we can observe the improvement in the cooling power, $\dot{Q}_1/\dot{\bar{Q}}_1$, for $\alpha = 0.8$ within the same region of parameters. This reveals that higher improvements in the cooling ability of the fridge are obtained as we approach the reversible point of operation for equal temperatures $\beta_3, \beta_2 \rightarrow \beta_1$, where $\dot{Q}_1$ vanishes. 
Nevertheless, we are particularly interested in regions where the cooling power is high [bottom region in Fig. \ref{fig:3}A], where good enhancements (about $1.45 \dot{\bar{Q}}_1$ for this set of parameters) can be obtained.

An alternative way of quantifying the performance of the fridge is by using the (inverse) effective temperature of qubit $1$ in the steady state, namely $\beta_1^\mathrm{eff}$ as given by Eq.~\eqref{eq:efftemp}.  Analogously, the effective temperatures of qubits $2$ and $3$ in the steady state can be used to illustrate the operation of the fridge.
Inside the cooling window, qubit $2$ has higher temperature than reservoir $2$ while qubits $1$ and $3$ being cooler than their respective reservoirs. 
This witnesses that heat flows spontaneously from reservoirs $1$ and $3$ into the thermal machine, and from the thermal machine into reservoir $2$. 

The collective transitions in the fridge produce a larger difference between the temperatures of the reservoirs and the machine qubits as $\alpha$ is increased. In particular in Fig. \ref{fig:3}C and D we show $\beta_1^{\mathrm{eff}}$ as a function of the reserovir temperatures. Fig. \ref{fig:3}C corresponds to the case $\alpha = 0$. Following the previous notation we denote $\bar{\beta}_1^\mathrm{eff}$ the inverse effective temperature reached for separate reservoirs. We notice that the plot is very similar to Fig.~\ref{fig:3}A, showing the same qualitative behavior. Indeed we conclude that lower effective temperatures are reached in qubit $1$ when the heat flowing from the coldest reservoir is increased. Nevertheless, some differences arise when looking at the enhancements in the effective inverse termperatures for non-zero $\alpha$, that is the ratio $\beta_1^\mathrm{eff}/ \bar{\beta}_1^{\mathrm{eff}}$ which is shown in Fig. \ref{fig:3}D,  where it can be seen that enhancements are in general small and peak in a region of parameters displaced from the one reaching maximum enhancements in cooling power [Fig. \ref{fig:3}B] (equal temperatures). 
In this case, the maximum enhancement is obtained for moderate (inverse) temperatures of reservoir $3$, i.e. $\beta_3 \simeq 0.3 \beta_2$. These differences arise from the fact that the additional transitions in the model for $\alpha >0$, allow heat to flow from the cold reservoir into the machine other than through qubit $1$. Therefore, the effective temperature of qubit $1$ may no longer completely determine the magnitude of $\dot{Q}_1$, as it was the case for the original model with $\alpha=0$.

The improvements in the cooling power of the refrigerator may be intuitively understood by turning back to the cooling mechanism of the model, as introduced in Sec. \ref{sec:setup}. 
In the case $\alpha = 0$ (original model), a cycle extracting a quantum of energy $E_1$ from reservoir at $\beta_1$ and $E_3$ from reservoir at $\beta_3$ and dissipating $E_2$ into reservoir at $\beta_2$ looks like
\begin{equation} \label{eq:cycle}
 \ket{010} {\color{blue} \rightarrow} \ket{110} {\color{red} \rightarrow} \ket{111} {\color{orange} \rightarrow} \ket{101} \rightarrow \ket{010},
\end{equation}
where the colors indicate which thermal reservoir mediates the transitions and we used black for the interaction Hamiltonian, $H_\mathrm{int}$ in Eq.~\eqref{eq:int}. In fact, there are exactly $6$ equivalent cycles of length $4$ 
starting in $\ket{010}$ and arriving with the help of the reservoirs to state $\ket{101}$, to get finally reset back to $\ket{010}$ by means of $H_\mathrm{int}$. Nevertheless, as soon as common reservoirs are considered each of the cycles 
split into two, now allowing for a shorter path. For example for the cycle \eqref{eq:cycle} we will have
\begin{equation} \label{eq:cycle2}
 \ket{010} {\color{blue} \rightarrow} \ket{110} {\color{red} \rightarrow} \ket{111} {\color{orange} \rightarrow}  \ket{010},
\end{equation}
since now the latter transition is allowed by the common reservoir. Therefore the previous $6$ cycles of length $4$ are combined for $\alpha > 0$ with $6$ new cycles of length $3$ as the one in~\eqref{eq:cycle2}, which may accomplish the task of cooling more quickly.

\subsection{Efficiency and entropy production}
\label{sec:efficiency}

\begin{figure*}[t!]
\includegraphics[width= 0.85 \linewidth]{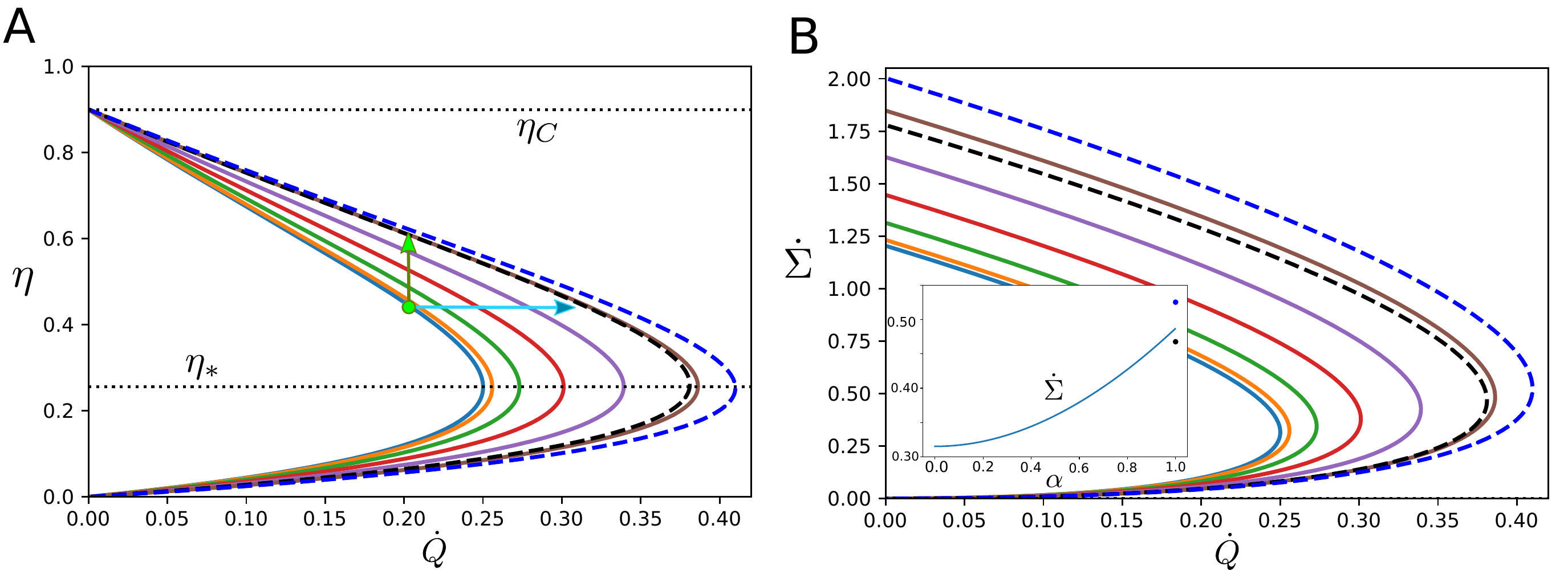}
\caption{Parametric plot of cooling power versus the COP $\eta$ (A) and versus entropy production rate $\dot{\Sigma}$ (B) for $\beta_2 = 0.5 \beta_1$ and $\beta_3 = 0.05 \beta_1$. As in the previous plot, solid lines correspond to values $\alpha = \{0, 0.2, 0.4, 0.6, 0.8, 0.99 \}$ (from left to right) 
while dashed lines correspond to our main two choices for the initial state of the refrigerator when $\alpha = 1$. The dotted lines in (A) correspond to the Carnot COP, $\eta_C = 0.9$, and COP at maximum cooling power $\eta_\ast \simeq 0.253$ for $\alpha \in [0,1)$. The inset in (B) shows the entropy production rate at maximum cooling power as a function of $\alpha$. Again the two circles at $\alpha = 1$ correspond to the two different initial states. Rest of parameters are $\beta_2 = 0.5 \beta_1$, $\beta_3 = 0.05 \beta_1$, and $g= 0.005$. Cooling power is given in $\gamma_0$ units.  \label{fig:4}}
\end{figure*}

As we previously mentioned, the enhancements in the cooling power of the machine due to the presence of common reservoirs come essentially at no efficiency cost, since $\eta$ does not change significantly when $\alpha$ is varied. 
Now we explore in more detail the tradeoff between cooling power and efficiency by constructing a parametric plot relative to both quantities [Fig. \ref{fig:4}A]. As before, solid curves with different colors correspond 
to different values of $\alpha \in [0, 1)$, while dashed lines correspond to the case $\alpha = 1$ with initial condition as in the previous figure: a product of thermal states and a state orthogonal to the dark one.

We find that increasing $\alpha$ always implies a greater cooling power $\dot{Q}_1$ for any fixed value of the COP, $\eta$.  Moreover the cooling power vanishes when approaching Carnot COP, $\eta_C$, which corresponds to the case of zero entropy production as expected for endoreversible refrigerators~\cite{Adesso}. The relation between entropy production and cooling power is highlighted in Fig \ref{fig:4}B. There, high entropy productions (top part of the plot) correspond to the regime where 
low efficiencies are obtained [bottom part of Fig. \ref{fig:4}A] and viceversa. Although the presence of collective dissipation increases the irreversibility in highly inefficient regimes, we see that it can also lead to a reduction 
of the entropy production rate below maximum power conditions. Of particular interest is the entropy production rate at maximum power, which we show in the inset figure as a function of $\alpha$. For such conditions, it becomes clear that the presence of the common reservoir push the steady state slightly more away from equilibrium, while leading to a beneficial effect in the cooling power, which almost doubles its magnitude. 

Turning back to Fig \ref{fig:4}A we remark that introducing common reservoirs  not only allows us to improve the cooling power for a fixed value of the efficiency (blue arrow), but it can be also used to improve efficiency while maintaining a 
fixed cooling power output (green arrow). This two possibilities can be combined as well to fit specific requirements on the operation of the refrigerator.

\subsection{Incoherent correlated (ic) dissipation}
\label{sec:classical}
One of the major goals of quantum thermodynamics is to identify the thermodynamic consequences of quantum effects. 
As we have seen before, the structure of the autonomous refrigerator studied here, combined with the presence of the three common reservoirs, leads to environmentally-induced coherence in its steady state. Here we discuss the role played by this environmental coherence in the enhancements reported above by providing a comparison to a correlated but incoherent model of dissipation.

The collective dissipation induced by three common baths as in Eqs.~\eqref{eq:master}-\eqref{eq:systemops} entails different effects: 
(i) the coherent sum of different one- and two-spin terms in each jump operator (\ref{eq:systemops}) due to the resonance of the respective processes, 
and (ii) the flip of spins pairs described by (three) quadratic spin terms. 
Let us now explore what happens when the processes generating a standard single spin flip are independent of the ones achieving paired spin transitions.
This dissipative model corresponds to relaxing the condition (i) still maintaining each nonlocal (two-spin) process (ii), that is nonlinear collective effects. 
In other words, each spin suffers usual local dissipation and also {\it independent} two-spin (thus non-local) dissipation. 
Due to the persistence of correlations entailed by two-spin flips, we refer to this case as incoherent correlated (ic) dissipation to be compared to the coherent and fully correlated dissipation model considered before. 

The corresponding model is obtained by replacing the environment-induced collective transitions $s_i$ [Eq.~\eqref{eq:systemops}] in the master equation~\eqref{eq:master}, by incoherent uncorrelated transitions.
In such case the Lindbladians in Eq.~\eqref{eq:lindbladians} split into two terms. For instance the Lindbladian accounting for dissipative processes at frequency $E_1$ now reads
\begin{align} \label{eq:lindbladians-classical}
 \mathcal{L}_1(\rho_\mathrm{m}) &= \gamma_\downarrow^{1} \left(\sigma_1^{-} \rho_\mathrm{m} \sigma_1^{+} - \frac{1}{2}\{ \sigma_1^{+} \sigma_1^{-}, \rho_\mathrm{m} \} \right)  \\
& ~+ \gamma_\uparrow^{1} \left(\sigma_1^{+} \rho_\mathrm{m} \sigma_1^{-} - \frac{1}{2}\{ \sigma_1^{-} \sigma_1^{+}, \rho_\mathrm{m} \} \right) \nonumber \\
&+ \alpha^2 \gamma_\downarrow^{1} \left(\sigma_2^{-} \sigma_3^{+} \rho_\mathrm{m} \sigma_2^{+} \sigma_3^{-} - \frac{1}{2}\{ \sigma_2^{+} \sigma_2^{-} \sigma_3^{-} \sigma_3^{+}, \rho_\mathrm{m} \} \right) \nonumber \\
& ~+ \alpha^2 \gamma_\uparrow^{1} \left(\sigma_2^{+} \sigma_3^{-} \rho_\mathrm{m} \sigma_2^{-} \sigma_3^{+} - \frac{1}{2}\{ s_2^{-} s_2^{+} \sigma_3^{+} \sigma_3^{-}, \rho_\mathrm{m} \} \right), \nonumber
 \end{align}
and analogously for terms $2$ and $3$. That is, the main difference with respect to the model introduced in Sec. \ref{sec:setup} is that now the cross-terms appearing in Eq.~\eqref{eq:lindbladians} are absent. 
Notably, this is exactly equivalent to removing all terms in the equations of motion  coupling populations of energy levels with the real part of the coherence $c_\mathcal{R}$ in Eqs.~\eqref{eq:dynamics1}-\eqref{eq:dynamics3}(see also Appendix~\ref{appB}). 
This means that all the terms in the equations depending on single powers of $\alpha$ disappear (but not those depending on $\alpha^2$). 
In particular, the heat currents from the three reservoirs $\dot{Q}_j$ defined in Eqs. \eqref{eq:currents}-\eqref{eq:currents3} will just drop the terms proportional to $2 \alpha c_\mathcal{R}^\pi$. 
We denote the heat currents in the incoherent correlated dissipation model by $\dot{Q}_j^{\mathrm{ic}}$, and in particular the cooling power by $\dot{Q}_1^{\mathrm{ic}}$.

\begin{figure*}[t!]
\includegraphics[width= 0.55 \linewidth]{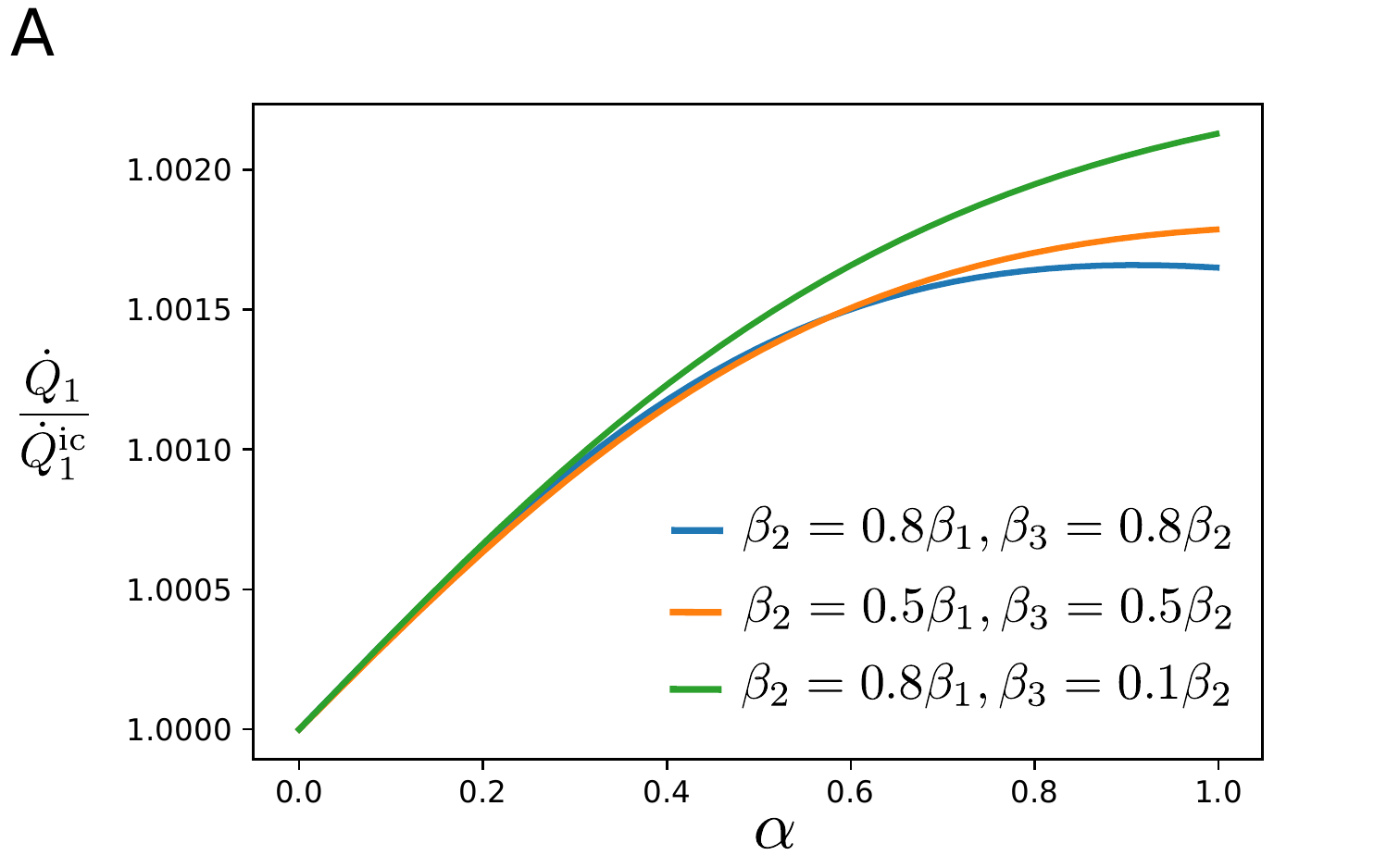}
\includegraphics[width= \linewidth]{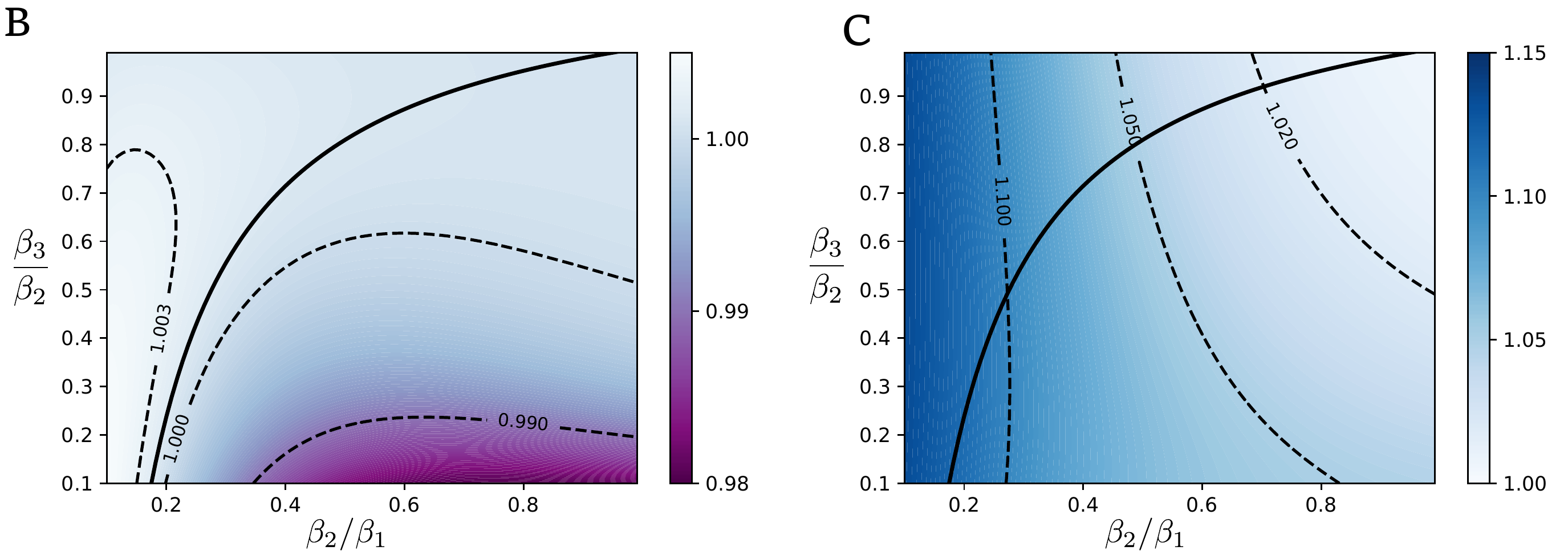}
\caption{Ratio between the cooling power of the fridge in the coherent and incoherent dissipation models $\dot{Q}_1/\dot{Q}_1^{\mathrm{ic}}$: (A) as a function of $\alpha \in [0, 1)$ for different choices of parameters. (B)-(C) for $\alpha = 1$ as a function of the inverse temperaures of the reservoirs for two different initial conditions for the refrigerator. In the case of an initial product thermal state (B) improvements in the collective coherent model are negligible while they become more important in the case of an initial state completely orthogonal to the dark state $\ket{\psi_D}$ (C).} \label{fig:5} 
\end{figure*}


Our results show that for most choices of parameter in the range $\alpha \in [0, 1)$, the models with (\ref{eq:systemops}) and (\ref{eq:lindbladians-classical}) are very similar, even though the collective-coherent case shows slight improvements. 
Nevertheless, a different situation arises in the case in which $\alpha = 1$. At this point the steady state predicted by Eq.~(\ref{eq:master}) suffers a finite-size phase transition responsible for the decoherence-free subspace (\ref{eq:dark}) 
which is absent here. In such case, the differences between the performance of the two models become appreciable, and sensible to the initial state of the refrigerator.
This is because for $\alpha = 1$ the dynamics becomes completely independent of the dark state $\ket{\phi_D}$ in Eq.~\eqref{eq:dark} in the coherent dissipation case. 
Therefore the refrigeration mechanism becomes also independent of $\ket{\psi_D}$, so that if the machine has some probability to be in the dark state $p_D$, then with 
this same probability it will not refrigerate. Consequently, initial states with $p_D = 0$ will perform better, more quickly than in other cases, since they will also have 
$p_D = 0$ at steady-state conditions (see Fig. \ref{fig:2}). As we see, in this case the protection of the dark state against dissipation and decoherence is useful when 
this state is not populated, contrary to what usually  happens in quantum information applications~\cite{Lidar}.

Figure~\ref{fig:5} summarizes our results for the ratio between the cooling powers of the fridge for quantum and classical models, $\dot{Q}_1/\dot{Q}_1^{\mathrm{ic}}$. 
In Fig. \ref{fig:5}A this ratio is plotted as a function of $\alpha \in [0,1)$ for different values of the parameters as indicated in the legend. 
We obtain that the difference in the performance between coherent and incoherent dissipation models is very small for $\alpha < 1$, below $\dot{Q}_1 \sim 1.002~\dot{Q}_1^\mathrm{ic}$. 
The case $\alpha=1$ is instead shown in Fig.~\ref{fig:5}B and C where we plot the same ratio $\dot{Q}_1/\dot{Q}_1^{\mathrm{ic}}$ as a function of the inverse temperatures of the reservoirs. 
In Fig~\ref{fig:5}C, where $\rho_\mathrm{ini} \ket{\psi_D} = 0$, i.e. $p_D = 0$, we find a range of inverse temperatures inside the cooling window for which the coherent model performs appreciably better, 
that is $\dot{Q}_1 \sim 1.1~\dot{Q}_1^\mathrm{ic}$. This enhancement is due to the choice of the initial population of the decoherence free subspace as explained before and is therefore of quantum origin. 
The advantage instead dissapears for other choices like a product initial state in the fridge $\rho_\mathrm{ini} = \rho_1^{\beta_1} \otimes \rho_2^{\beta_2} \otimes \rho_3^{\beta_3}$. 
As we can see from Fig.~\ref{fig:5}B for a large area inside the cooling window $\dot{Q}_1/\dot{Q}_1^{\mathrm{ic}} < 1$, that is, the incoherent correlated dissipation model performs slightly better than the coherent one.


\section{Conclusions}
\label{sec:conclusions}

Correlated transitions entailed by collective dissipation can significantly improve the performance of small quantum autonomous refrigerators. 
The enhancements are manifested in a boost of the cooling power of the fridge as well as in lower reachable  temperatures without compromising the efficiency.
We have found that the biggest improvements in the performance are achieved for 
the case in which dissipative effects are induced by three fully common reservoirs ($\alpha = 1$) leading to coherent one and two spin transitions. 
In this situation a decoherence free subspace emerges, which can significantly alter the dynamics of the refrigeration process.
On the other hand, for $\alpha\neq 1$, coherent effects in the dissipative dynamics are not crucial and similar enhancements can be also achieved in a incoherent partial correlated model, where one and two spin processes occurr independently. 
This indicates that common noise may be also considered for enhancing the performance of small autonomous classical motors, like thermoelectric devices~\cite{thermoelectric} or biomolecular systems~\cite{biochemical} described within the framework of stochastic thermodynamics~\cite{stochastic}. 
Nevertheless we notice that although the magnitude of the quantum coherent enhancements compared with their incoherent counterpart is rather modest in this fridge model, we expect that the same mechanism might be enforced in many-body configurations~\cite{Saro,Kurizki,Raam} where this effect may be greatly amplified.

Furthermore, we find that whenever $\alpha$ approaches $1$, the effect of the interaction Hamiltonian between the three fridge qubits in Eq.~\eqref{eq:int} becomes superflous for the function of the fridge. 
This can be already understood from the availability of cycles like the one pointed in \eqref{eq:cycle2}. Therefore, using the collective couplings to the environment reported here [Eq.~\eqref{eq:systemops}] becomes an alternative way to realize the fridge wihtout the need of the three-body interaction in Eq.~\eqref{eq:int}.
The consistency of the local approach employed in the description of the refrigerator has been checked by computing the higher order contributions to the heat currents neglected in Eqs.~\eqref{eq:currents}. 
We obtain that the magnitude of these corrections are always of order lower than $10^{-3} |\dot{Q}_i|$ for the range of parameters considered here.

The possibility of common bath leading to collective dissipation has been reported in different experimental platforms as mentioned in Sect. I. 
More challenging is the implementation of two-spin flip terms. As recently proposed in Ref.~\cite{nori}, a pair of two-level atoms coupled to a single-mode resonator with cavity frequency twice the qubit transition, can actually display joint absorption. 
Superconducting artificial atoms are shown to be suitable for such processes to take place using superconducting transmission line resonators and placing the qubits at the extreme points of such lines~\cite{majer}.

In this work we restricted ourselves to symmetric spontaneous decay rates and the long time behavior of the quantum fridge. Going beyond these two assumptions may be an interesting direction for future studies on this model.
In particular it has been shown that the use of asymmetric decays in the case of separate reservoirs may help in the performance of the refrigerator~\cite{Popescu}. Our results show that this is also the case for $\alpha > 0$. However, a detailed characterization of the impact of asymmetric rates on the performance of the fridge may be desirable. Also, we notice that some recent works  explored the transient dynamics of the same model for the case of separate reservoirs~\cite{Coherence-assisted, Transient}. Interestingly, coherence in the transport subspace induce coherent oscillations of the machine qubits temperatures during the evolution, leading to lower transient temperatures compared with those obtained in the steady state regime. We expect that these effects might also be enhanced by the presence of the common environmental effects reported here. Moreover, it will be interesting to explore such possible enhancements in connection with other non-trivial dynamical features of common reservoirs, such as spontaneous transient synchronization \cite{GiorgiSync, ManzanoNet, GiorgiSpins}, 
the generation of boundary time-crystals \cite{Iemini}, improving laser cooling~\cite{Morigi}, or boosting the charging process of quantum batteries~\cite{GooldBat, Campisi2}.

\begin{acknowledgments}
RZ and GLG acknowledge support from MINECO/AEI/FEDER through projects EPheQuCS FIS2016-78010-P, CSIC Research Platform PTI-001, the Maria de Maeztu Program for Units of Excellence in R$\&$D (MDM-2017-0711), and funding from CAIB postdoctoral program.
\end{acknowledgments}

\appendix

\section{Derivation of the master equation}
\label{appA}

The environment is made up of three reservoirs, each of them coupled to the three qubits.  The system-bath interaction is such that the emission (or absorption) of one excitation in the environment can be due either to a standard single spin flip or to a double spin transition \cite{nori}. Assuming a parameter $\alpha$ as the ratio between the probabilities of the two-atom and the single-atom processes, the interaction Hamiltonian can be written as
\begin{equation}\label{totalME}
H_I=\left[\sum_{i=1}^3 \sigma_i^x +\alpha {\sum_{ i>j}} \sigma_i^x\sigma_j^x \right] \sum_{l=1}^3 B_l,
\end{equation}
where the three bath operators are
\begin{equation}
B_l=\sum_k \lambda_k^{(l)}(a^{\dag (l)}_k+a^{( l)}_k),
\end{equation}
where $[a_k^{(l)}, a_{k'}^{\dagger (l)}]= \delta_{k, k^\prime}$ are the bosonic ladder operators of modes $k$ and $k^\prime$ of reservoir $l$. The baths' Hamiltonian is given by 
\begin{equation} \label{eq:bath}
 H_{R_l} = \sum_k \hbar \Omega_k a_k^{\dagger (l)} a_k^{(l)}.
\end{equation}

The coupling of the three baths to the system is assumed to be small enough such that the  linewidth  is much smaller than the system energies $E_i$. The $l$th environment is characterized by a spectral density peaked around $E_l$,  and with approximately no modes around the other two frequencies. Due to the smallness of the coupling,  each reservoir can be assumed to have an almost Ohmic behavior about the relevant frequency \cite{vasile}.  Thus, the densities of states obey $J_l(E_i)\simeq \gamma_0^{(i)}\delta_{l,i}/2 \pi $. These assumptions imply that the three environments are not correlated to each other and  lead to a Born-Markov master equation where the dissipative part is the sum of three separate contributions, each of them due to one of the three reservoirs. Furthermore,  in the limit of $g$ small with respect to the natural energies, a local approach to the master equation can be adopted,  which allows one to compute the Lindbladian using local operators \cite{RivasMEQ, HoferMEQ, GonzalezMEQ, 
PlenioMEQ,CattaneoMEQ}.

Due to the resonance condition $E_1+E_3=E_2$, for each of the three frequencies both one-atom and two-atom processes can take place within the bandwidth of the respective bath: a photon with energy $E_1$ can induce a single flip in the atom $1$ associated to the system-bath operator $a_1^\dag \sigma_1^-$ or a double flip where  the  operator is given as $a_1^\dag \sigma_2^-\sigma_3^+$. (together with the Hermitian conjugate processes).  Equivalently a photon with energy $E_2$ can induce processes as  $a_2^\dag \sigma_2^-$ or $a_2^\dag \sigma_1^-\sigma_3^-$, while a photon with energy $E_3$ can produce the transitions $a_3^\dag \sigma_3^-$ or $a_3^\dag \sigma_1^+\sigma_2^-$. The coherence of the two classes of processes (one-atom and two-atom flips) is guaranteed by the fact that any of the three environments influence all the atoms [see Eq.~(\ref{totalME})]. 
This is taken into account writing the jump operators according to Eq.~(\ref{eq:systemops}), which determine the form of the Lindbladian superoperators of Eq.~(\ref{eq:lindbladians}) that, once the rotating-wave approximation has been performed, give the master equation~(\ref{eq:master}).
The assumption of independent baths for one- and two spin-processes described in Sec.\ref{sec:classical}  leads in a similar way to a master equation with Lindbladian~(\ref{eq:lindbladians-classical}).

\section{Populations and coherence dynamics}
\label{appB}

In this appendix we report the full set of coupled differential equations for the populations of the 8 energy levels of the fridge and the real and imaginary part of the coherence between levels $\ket{010}$ and $\ket{101}$ as derived from the Lindblad master equation~\eqref{eq:master}.
\begin{widetext} 
\begin{align} \label{eq:ap1}
 \dot{p}_{000} =&~ \gamma_\downarrow^{(1)} p_{100} + \gamma_\downarrow^{(2)} p_{010} + \gamma_\downarrow^{(3)} p_{001} - p_{000}\left( \gamma_\uparrow^{(1)} + \gamma_\uparrow^{(2)} + \gamma_\uparrow^{(3)} \right),  \\
 \dot{p}_{001} =&~ \gamma_\downarrow^{(1)} p_{101} + \gamma_\downarrow^{(2)} p_{011} + \gamma_\uparrow^{(3)} p_{000} - p_{001}\left( \gamma_\uparrow^{(1)} + \gamma_\uparrow^{(2)} + \gamma_\downarrow^{(3)} \right),  \\
 \dot{p}_{100} =&~ \gamma_\uparrow^{(1)} p_{000} + \gamma_\downarrow^{(2)} p_{110} + \gamma_\downarrow^{(3)} \left( \alpha^2 p_{010} + p_{101} + 2 \alpha c_\mathcal{R} \right) - p_{100}\left( \gamma_\downarrow^{(1)} + \gamma_\uparrow^{(2)} + (\alpha^2 +1) \gamma_\uparrow^{(3)} \right),  \\
 \dot{p}_{101} =&~ \gamma_\uparrow^{(2)} p_{000} + \alpha^2 \gamma_\uparrow^{(1)} p_{001} + \alpha^2 \gamma_\uparrow^{(3)}p_{100}  + \gamma_\downarrow^{(3)} p_{011} + \gamma_\downarrow^{(1)} p_{110} + \alpha^2 \gamma_\downarrow^{(2)} p_{111} - 2i g c_\mathcal{I}  \\
                &~  - \alpha \sum_i (\gamma_\uparrow^{(i)} + \gamma_\downarrow^{(i)}) c_\mathcal{R} -[\gamma^{(1)}_\uparrow + \gamma^{(2)}_\downarrow + \gamma^{(3)}_\uparrow + \alpha^2 (\gamma^{(1)}_\downarrow + \gamma^{(2)}_\uparrow + \gamma^{(3)}_\downarrow)] p_{010},  \\
 \dot{p}_{010} =&~ \alpha^2 \gamma_\uparrow^{(2)} p_{000} + \gamma_\uparrow^{(1)} p_{001} + \gamma_\uparrow^{(3)}p_{100}  + \alpha^2 \gamma_\downarrow^{(3)} p_{011} + \alpha^2 \gamma_\downarrow^{(1)} p_{110} + \gamma_\downarrow^{(2)} p_{111} + 2i g c_\mathcal{I}  \\
                &~  - \alpha \sum_i (\gamma_\uparrow^{(i)} + \gamma_\downarrow^{(i)}) c_\mathcal{R} -[\alpha^2 (\gamma^{(1)}_\uparrow + \gamma^{(2)}_\downarrow + \gamma^{(3)}_\uparrow) + \gamma^{(1)}_\downarrow + \gamma^{(2)}_\uparrow + \gamma^{(3)}_\downarrow] p_{010},   \\
 \dot{c}_\mathrm{R} =&~ \alpha (\gamma^{(2)}_\uparrow p_{000} + \gamma^{(1)}_\uparrow p_{001} + \gamma^{(3)}    \uparrow p_{100} + \gamma^{(3)}_\downarrow p_{011} + \gamma^{(1)}_\downarrow p_{110} + \gamma^{(2)}_   \downarrow p_{111})  \\
                     &~ - \alpha \sum_i (\gamma^{(i)}_\uparrow + \gamma^{(i)}_\downarrow)(p_{010} + p_{101}) - (\alpha^2 +1)\sum_i (\gamma^{(i)}_\uparrow + \gamma^{(i)}_\downarrow), \\
 \dot{c}_\mathcal{I} =&~ i g (p_{101} - p_{010}), \\
 \dot{p}_{011} =&~ \gamma_\downarrow^{(1)} p_{111} + \gamma_\uparrow^{(2)} p_{001} + \gamma_\uparrow^{(3)} \left( \alpha^2 p_{101} + p_{010} + 2 \alpha c_\mathcal{R} \right) - p_{011}\left( \gamma_\uparrow^{(1)} + \gamma_\downarrow^{(2)} + (\alpha^2 +1) \gamma_\downarrow^{(3)} \right),  \\
 \dot{p}_{110} =&~ \gamma_\uparrow^{(1)} p_{010} + \gamma_\uparrow^{(2)} p_{100} + \gamma_\downarrow^{(3)} p_{111} - p_{110}\left( \gamma_\downarrow^{(1)} + \gamma_\downarrow^{(2)} + \gamma_\uparrow^{(3)} \right), \\
 \dot{p}_{111} =&~ \gamma_\uparrow^{(1)} p_{011} + \gamma_\uparrow^{(2)} p_{101} + \gamma_\uparrow^{(3)} p_{110} - p_{111}\left( \gamma_\downarrow^{(1)} + \gamma_\downarrow^{(2)} + \gamma_\downarrow^{(3)} \right), \label{eq:ap2}
\end{align}
\end{widetext}
Moreover, following the master equation~\eqref{eq:master}, we find that the other non-diagonal elements of the fridge density operator suffer exponential decay and become zero in the long time run. Therefore they become completely irrelevant for the steady state operation of the fridge. 

The above set of equations \eqref{eq:ap1}-\eqref{eq:ap2} can be rewritten in matrix form as $\dot{\bold{p}} = \mathcal{W} \bold{p}$, with  $\bold{p} = (p_{000}, ..., p_{111}, c_\mathcal{R}, c_\mathcal{I})^T$. We obtain the relevant elements of the steady state density operator, $\bold{\pi} = (\pi_{000}, ..., \pi_{111}, c^\pi_\mathcal{R}, c^\pi_\mathcal{I})^T$, by obtaining the kernel of the dynamical matrix $\mathcal{W}$, that is, $\mathcal{W} \bold{\pi} = 0$.

\end{document}